\documentstyle[12pt]{article}
\begin{document}

\baselineskip=24pt plus 2pt
\vskip -5cm
\hfill\hbox{NCKU-HEP/95-01}
\vskip -0.2cm
\hfill\hbox{28 January 1995}
\vskip 1cm
\begin{center}
{\large \bf  Effective Potential of Nambu--Jona-Lasinio Model \\
in Differential Regularization}\\
\vspace{20mm}
Yaw-Hwang Chen \\
and\\
Su-Long Nyeo\\
\vspace{5mm}

Department of Physics, National Cheng Kung University \\
Tainan, Taiwan 701, Republic of China \\

\vspace{10mm}
\end{center}
\begin{center}
{\bf ABSTRACT}
\end{center}
The method of differential regularization is applied to calculate explicitly
the one-loop effective potential of a bosonized Nambu--Jona-Lasinio model
consisting of scalar and pseudoscalar fields.  The regularization scheme
independent relation for the $\sigma$ mass sum rule is obtained.
This method can be readily applied to extended NJL models with gauge fields.

\newpage
\noindent
{\large \bf 1. Introduction}

A major goal of low-energy hadron physics consists in obtaining effective
Lagrangians, which incorporate certain features of quantum chromodynamics
that are relevant for describing the properties of hadrons at low and
intermediate energies.  Among many models, the Nambu--Jona-Lasinio (NJL)
model [1] seems to be the simplest pure quark theory, which yields
dynamical symmetry breaking and hence a non-vanishing value for the
quark condensate.  But to make the perturbation series non-divergent,
we need to introduce an ultra-violet cut-off as a regularization.
Many regularization methods have been employed [2] and it was found that
the one-loop effective potentials in various methods can vary substantially
with respect to the change of the renormalization scale.  In particular,
one should know that not all regularization methods are suitable to
NJL models.  Moreover, higher-loop corrections are also needed to
reduce the sensitivity of the renormalization scheme dependence
of perturbative results.

In this paper, we shall use a space-time regularization method known
as differential regularization (DR) [3] to calculate the one-loop effective
potential of a simple NJL model.  This method has been shown to be useful
to study the quantum corrections in chiral theories.  By using DR method,
we can determine from the effective potential systematically and unambiguously.
 It is hope that such a space-time
approach can provide a systematic study for extended NJL models [4,5]
in which gauge fields are also considered.  We organize the paper as follows.
In section 2, we introduce the NJL model and the calculational
procedure of DR method.  In section 3, we calculate explicitly the one-loop
effective action of the NJL model, from which we obtain the one-loop effective
potential and determine the $\sigma$ mass sum rule and the quark condensate.

\vskip 1cm
\noindent
{\large \bf 2. The NJL Model and Differential Regularization}

We consider the NJL Lagrangian
for scalar and pseudoscalar couplings in the SU(2) sector
\begin{equation}
L_{NJL} = \bar{q} i \gamma^{\mu} \partial_{\mu} q +
          {G\over 2}\left[(\bar{q} q)^2 + (\bar{q} i
          \vec{\tau}\gamma_5 q)^2\right] - M_{0} \bar{q} q\,,
\end{equation}
where $q$ stands for the isospin SU(2) quark field with 3 colors
and current mass $M_{0}=m_{u}=m_{d}$. G is the coupling constant
with mass dimensions $(mass)^{-2}$ in $4$ dimensions.  ${\vec\tau}$ are
the Pauli matrices satisfying ${\rm Tr}_\tau[\tau^i\tau^j] = 2\delta^{ij}$.
Because of the four-fermion coupling, this Lagrangian is not renormalizable,
and for calculating quantum corrections,
an ultra-violet momentum cut-off must be introduced.

The quantized theory can be written in terms of a
generating functional, which, in the absence of external sources, reads
\begin{equation}
Z_{NJL} = \int D\bar{q} Dq \exp\left[i\int d^4x L_{NJL}(x)\right]\,\,.
\end{equation}
In order to use the model to describe the low-energy properties with the
manifest low-energy modes, we bosonize the model with scalar and pseudoscalar
fields $\sigma$ and $\vec{\pi}$. To do this, we multiply (2) by following
gaussian functional
\begin{equation}
1=\int D\sigma d\vec{\pi} \exp\left[-i\int d^4 x {\mu^2\over 2}
  \left((\sigma + {g\over \mu^2} (\bar{q} q - {M_{0}\over G}))^2 +
  (\vec{\pi} + {g\over{\mu^2}} \bar{q} i \vec{\tau}
  \gamma_{5} q)^2\right)\right]\,,
\end{equation}
which transforms the four-fermion interaction into a Yukawa-like coupling;
we obtain
\begin{equation}
{\cal Z}_{NJL} = \int D\bar{q} Dq D\sigma D\vec{\pi}
              \exp\left[i\int d^4x {\cal L}_{NJL}(x)\right]\,,
\end{equation}
where
\begin{equation}
{\cal L}_{NJL} = \bar{q} [i\gamma^{\mu} \partial_{\mu} -
              g(\sigma + i\vec{\pi} \cdot
              \vec{\tau} \gamma_{5})]q -
              {\mu^2 \over 2} (\sigma^2 + \vec{\pi}^2)
              + {M_{0}\mu^2 \over g} \sigma \,\,,
\end{equation}
with $G=g^2 /\mu^2$.  In low-energy approximation, we
couple the $\sigma$ field and the pion fields to external sources.
The quark fields are not coupled to any external sources and can be
integrated over to give a functional determinant, whose evaluation
requires a regularization. The calculation can be carried out by using
the Feynman rules for the quark fields, such that the one-loop quantum
effects are considered by treating quarks as internal lines in the Feynman
diagrams.

In this method, it is necessary to determine its vacuum by calculating the
effective potential to at least one-loop order.  We shall employ differential
regularization and carry out the calculation in Euclidean space, so that
the functional integral reads
\begin{equation}
Z_{NJL}' = \int D\sigma D{\vec\pi}
\exp\left[-{\cal S}_{eff}(\sigma,{\vec\pi})\right]\,,
\end{equation}
where the effective action in the one-loop fermion approximation is
\begin{eqnarray}
{\cal S}_{eff} = -\ln{\rm Det}[-i\gamma^{\mu} \partial_{\mu} &+&
              g(\sigma + i\vec{\pi} \cdot
              \vec{\tau} \gamma_{5})]\nonumber\\
               &+& \int d^4x\left[
              {\mu^2 \over 2} (\sigma^2 + \vec{\pi}^2)
              - f_\pi m_\pi^2\sigma\right] \,\,.
\end{eqnarray}
Note that we have used the PCAC condition $M_0\mu^2/g = f_\pi m_\pi^2$,
where $f_\pi$ is the pion decay constant
$(93 MeV)$ and $m_\pi$ is the pion mass $(139 MeV)$.

The functional determinant can be calculated using the Feynman rules, which are
given as follows.  The massless quark propagator is given by
\begin{eqnarray}
\langle {q_a^i(x) \bar{q}_b^j(0)} \rangle
&\equiv& S_{ab}^{ij}(x) \nonumber\\
&=& - {i{\delta^{ij}\delta_{ab}}\over{2\pi^2}} {\gamma_{\mu}
         x^{\mu} \over x^4}\,,
\end{eqnarray}
with $i,j$ and $a,b$ being isospin and color indices, respectively.
The Feynman rule for the $\bar{q} q \sigma$ vertex is
$-g\delta^4(x_2 - x_1)\delta^4(x_3 - x_1)$, and that
for the $\bar{q} q \vec{\pi}$ vertex is $-ig\vec{\tau}\gamma_{5}
\delta^4(x_2 - x_1)\delta^4(x_3 - x_1)$\,.

In perturbative calculations, we encounter highly-singular terms of the form
\begin{equation}
{1\over (x^2)^n} \ln^m(\mu^2 x^2),~~n \geq 2\,,
m \geq 0\,,
\end{equation}
where $\mu$ is a mass parameter in the problem.
The essential idea of differential regularization method is
to define these highly-singular terms by
\begin{equation}
{1\over (x^2)^n}\ln^m(\mu^2x^2)
= \underbrace{\Box\Box\ldots\Box}_{n - 1}G(x^2)\,\,,
x^2 \neq 0\,,
\end{equation}
where $G(x^2)$ is a to-be-determined function
that has a well-defined Fourier transform and can depend
on $2(n-1)$ integration constants, which are the short-distance cut-offs.
In this paper, we encounter only the following two forms:
\begin{equation}
{1\over (x^2)^2} \equiv -{1\over4}\Box{\ln x^2M^2 \over x^2}\,\,,
x^2 \neq 0\,,
\end{equation}
\begin{equation}
{1\over (x^2)^3} \equiv -{1\over{32}}\Box\Box{\ln x^2M^2 \over x^2}\,\,,
x^2 \neq 0\,.
\end{equation}
where the mass parameter M is an integration constant. Note that we have
omitted other irrelevant integration constants for $x^2\neq 0$.

This regularization method has been used in $\phi^4$ theory and
QCD and was able to reproduce the well-known results obtained by other methods.
The advantage of DR method is that loop corrections in a
chiral theory can be calculated unambiguously.  We note also that
different methods can lead to different results for the one-loop effective
potential, which depends strongly on renormalization scheme.  Higher-loop
corrections can reduce the sensitivity of scheme dependence, but of course,
use of a regularization method requires special attention.

\vskip 1cm
\noindent
{\large \bf 3. The One-Loop Effective Action}

To obtain the one-loop effective action of the NJL model,
we need to evaluate the one-loop self-energies of the $\sigma$ field
and the pion fields, and the vertices of the fields.
The self-energy of the $\sigma$ field
with an internal quark loop is easily calculated and reads
\begin{eqnarray}
\Pi_\sigma(x,y) & = & -g^2{\rm Tr}\left[S(x-y)S(y-x)\right] \nonumber\\
                & = & -{6g^2 \over \pi^4} {1\over (x-y)^6}\nonumber\\
         & = & {{3g^2}\over{16\pi^4}}\Box\Box{\ln (x-y)^2 M^2\over (x-y)^2}\,,
          {(x - y)}^2 \neq 0\,,
\end{eqnarray}
where $\Box = \Box_{(x-y)}$ and we have used ${\rm Tr}_{\gamma}I = 4,
{\rm Tr}_cI = 3,$ and ${\rm Tr}_{\tau}I = 2$ for the spinor, color,
and isospin degrees of freedom, respectively.
The self-energies of the pion fields $\pi^i$ are
\begin{eqnarray}
\Pi^{ij}_\pi(x,y) & = & -
g^2{\rm Tr}\left[i\tau^i\gamma_5S(x-y)i\tau^j\gamma_5S(y-x)\right] \nonumber\\
                  & = &
-{6g^2\over \pi^4} {\delta^{ij}
                              \over (x-y)^6} \nonumber\\
                  & = & {3g^2\over 16\pi^4} \delta^{ij}
                             \Box \Box {\ln (x-y)^2 M^2 \over (x-y)^2}\,\,,
          {(x - y)}^2 \neq 0\,.
\end{eqnarray}

One can show that as a consequence of the trace properties of the Dirac
gamma matrices in 4 dimensions, the one-loop diagram with one external
$\sigma$ field and one external pion field vanishes. Other non-vanishing
one-loop 4-point vertex functions with external boson
fields are the following three diagrams: one has 4 $\sigma$ fields, one has
4 pion fields, and another has 2 $\sigma$ fields and 2 pions.
The one-loop 4-point vertex function with 4 external $\sigma$ fields
is given by
\begin{equation}
\Gamma_{\!\sigma\sigma\sigma\sigma}(x_1,x_2,x_3,x_4) = G(x_1,x_2,x_3,x_4)
+ {\rm permutations}\,,
\end{equation}
where $\Delta_{ij} = x_i - x_j$, and
\begin{eqnarray}
G(x_1,x_2,x_3,x_4)  & = &
-g^4{\rm Tr}\left[S(\Delta_{12})S(\Delta_{23})S(\Delta_{34})S(\Delta_{41})
\right]\nonumber\\
     & = &
{{6g^4}\over{(2\pi^2)^4}}
{\rm Tr}_\gamma\left[\gamma_\mu\gamma_\nu\gamma_\alpha\gamma_\beta\right]
V_{\mu\nu\alpha\beta}\nonumber\\
     & = &
{{3g^4}\over{2\pi^8}}
(V_{\mu\mu\nu\nu} - V_{\mu\nu\mu\nu} + V_{\mu\nu\nu\mu})\,.
\end{eqnarray}
In terms of $x = x_1 - x_2, y = x_2 - x_3, z = x_3 - x_4$, the
function $V_{\mu\nu\alpha\beta}$ reads
\begin{eqnarray}
V_{\mu\nu\alpha\beta} & = &
{{x_\mu y_\nu z_\alpha (x + y + z)_\beta}\over{x^4y^4z^4(x + y +
z)^4}}\nonumber\\
               & =  &
{1\over16}{{\partial}\over{\partial x_\mu}}\left({1\over{x^2}}\right)
{{\partial}\over{\partial y_\nu}}\left({1\over{y^2}}\right)
{{\partial}\over{\partial z_\alpha}}\left({1\over{z^2}}\right)
{{\partial}\over{\partial x_\beta}}\left({1\over{(x + y + z)^2}}\right)\,.
\end{eqnarray}
We are interested in the singular parts of the $V$ functions, which
are given by
\begin{eqnarray}
V_{\mu\mu\nu\nu} &\buildrel sing. \over = & {1\over4}\pi^2\delta^4(z)\left[
{{\partial}\over{\partial x}}\left({1\over x^2}\right)\cdot
{{\partial}\over{\partial y}}\left({1\over y^2}\right)\right]
{1\over{(x + y)^2}}   \nonumber\\
&\buildrel sing. \over = & -{1\over4}\pi^2\delta^4(z) 4\pi^2\delta^4(x+y)
{1\over x^4} \nonumber\\
& = & {1\over4}\pi^4\delta^4(z) \delta^4(x+y) \Box {\ln x^2 M^2 \over x^2}\,,
x^2 \neq 0\,,\\
%
V_{\mu\nu\mu\nu}
&\buildrel sing. \over = &
{1\over4}\pi^4\delta^4(y) \delta^4(x+z) \Box {\ln z^2 M^2 \over z^2}\,,
z^2 \neq 0\,,\\
%
V_{\mu\nu\nu\mu}
&\buildrel sing. \over = &
{1\over4}\pi^4\delta^4(x) \delta^4(y+z) \Box {\ln y^2 M^2 \over y^2}\,,
y^2 \neq 0\,.
\end{eqnarray}
Hence
\begin{eqnarray}
G(x_1,x_2,x_3,x_4)  & = &
{3g^4\over8\pi^4}\left[\delta^4(\Delta_{34})\delta^4(\Delta_{13})
\Box{{\ln\Delta_{12}^2M^2}\over{\Delta_{12}^2}}\right.\nonumber\\
&-&\delta^4(\Delta_{23})\delta^4(\Delta_{14})
\Box{{\ln\Delta_{34}^2M^2}\over{\Delta_{34}^2}}\nonumber\\
&+&\left.\delta^4(\Delta_{12})\delta^4(\Delta_{24})
\Box{{\ln\Delta_{23}^2M^2}\over{\Delta_{23}^2}}\right]\,.
\end{eqnarray}
The 4-point vertex function with 4 external pion fields
can be calculated in the same manner and is given by
\begin{equation}
\Gamma^{(i\,j\,k\,l)}_{\pi\pi\pi\pi}(x_1,x_2,x_3,x_4)  =
G^{\,i\,j\,k\,l}(x_1,x_2,x_3,x_4) + {\rm permutations }\,,
\end{equation}
where
\begin{eqnarray}
G^{\,i\,j\,k\,l}(x_1,x_2,x_3,x_4)  & = &
-g^4{\rm Tr}\left[i\tau^i\gamma_5S(\Delta_{12})i\tau^j\gamma_5S(\Delta_{23})
\right.\times\nonumber\\
&   &
\left.i\tau^k\gamma_5S(\Delta_{34})i\tau^l\gamma_5S(\Delta_{41})
\right] \nonumber\\
& = &
{{3g^4}\over{4\pi^8}}
(V_{\mu\mu\nu\nu} - V_{\mu\nu\mu\nu} + V_{\mu\nu\nu\mu})
{\rm Tr}_{\tau}\left[\tau^i \tau^j \tau^k \tau^l\right]
\nonumber\\
& = & {1\over2}G(x_1,x_2,x_3,x_4)
{\rm Tr}_\tau\left[\tau^i \tau^j \tau^k \tau^l\right] \,,
\end{eqnarray}
where
${\rm Tr}_\tau\left[\tau^i\tau^j\tau^k\tau^l\right] = 2\delta^{ij}\delta^{kl}
- 2\delta^{ik}\delta^{jl} + 2\delta^{il}\delta^{jk}\,$.
Finally, we list the 4-point vertex functions with mixed external fields:
\begin{equation}
\Gamma^{\,i\,j\,\,\cdot\,\,\cdot}_{\pi\pi\sigma\sigma}(x_1,x_2,x_3,
x_4) = G(x_1,x_2,x_3,x_4)
\delta^{ij}\,,
\end{equation}
\begin{equation}
\Gamma^{\,i\,\cdot\,j\,\cdot}_{\pi\sigma\pi\sigma}(x_1,x_2,x_3,
x_4) = -G(x_1,x_2,x_3,x_4)
\delta^{ij}\,,
\end{equation}
\begin{equation}
\Gamma^{\,i\,\cdot\,\cdot\,j}_{\pi\sigma\sigma\pi}(x_1,x_2,x_3,
x_4) = G(x_1,x_2,x_3,x_4)
\delta^{ij}\,,
\end{equation}
\begin{equation}
\Gamma^{\,\cdot\,\,i\,j\,\cdot}_{\sigma\pi\pi\sigma}(x_1,x_2,x_3,
x_4) = G(x_1,x_2,x_3,x_4)
\delta^{ij}\,,
\end{equation}
\begin{equation}
\Gamma^{\,\cdot\,i\,\cdot\,j}_{\sigma\pi\sigma\pi}(x_1,x_2,x_3,
x_4) = - G(x_1,x_2,x_3,x_4)
\delta^{ij}\,,
\end{equation}
\begin{equation}
\Gamma^{\,\cdot\,\cdot\,i\,j}_{\sigma\sigma\pi\pi}(x_1,x_2,x_3,
x_4) = G(x_1,x_2,x_3,x_4)
\delta^{ij}\,,
\end{equation}
where $G(x_1,x_2,x_3,x_4)$ is given by (21).

Higher-point vertex functions either vanish or do not provide logarithmic
contributions and are omitted here.

Summing all the logarithmic contributions, we get
(up to quartic in the fields) the one-loop effective action
before renormalization
\begin{eqnarray}
\Gamma^{(1)}[\sigma,\vec{\pi}]
& = & {-3g^2\over 32\pi^4}\int d^4 x d^4 y (\Box\Box
{\ln (x - y)^2 M^2 \over (x - y)^2 }) \left[\sigma(x)\sigma(y)
+ \vec{\pi}(x)\cdot\vec{\pi}(y)\right]
\nonumber\\
&   & - {3g^4\over 32\pi^4}
\int d^4 x d^4 y (\Box{\ln (x - y)^2 M^2 \over (x - y)^2 })
\left[2\sigma^3(x)\sigma(y) - \sigma^2(x)\sigma^2(y)\right. \nonumber\\
&   &
{\hskip 4cm} + 2{\vec\pi}^2(x){\vec\pi}(x)\cdot{\vec\pi}(y) -
{\vec\pi}^2(x){\vec\pi}^2(y)\nonumber\\
&   &
{\hskip 4cm} - 2{\vec\pi}^2(x)\sigma^2(y) +
2{\vec\pi}^2(x)\sigma(x)\sigma(y)\nonumber\\
&   &
{\hskip 5cm} + \left.2{\vec\pi}(x)\cdot{\vec\pi}(y)\sigma^2(x)\right]\,.
\end{eqnarray}
Now using the fact that $\Box{1\over x^2} = -4\pi^2\delta^4(x)$, we get
\begin{eqnarray}
{\partial\Gamma^{(1)}[\sigma,\vec{\pi}] \over \partial t} & = &
{3g^2 \over 8\pi^2}\int d^4 x \left[\left(\partial_\mu\sigma(x)\right)^2
+ \left(\partial_\mu{\vec\pi}(x)\right)^2\right]
\nonumber\\
&   & +{3g^4 \over 8\pi^2}\int d^4 x \left(\sigma^2(x)
+ \vec{\pi}^2(x)\right)^2\,,
\end{eqnarray}
where $t = \ln\left(M^2/\Lambda^2\right)$ with $\Lambda$ being
the renormalization
scale.  From the general form of the effective action in Euclidean space
\begin{equation}
\Gamma(\sigma,\vec{\pi}) = \int d^4 x \left[+V_{eff}(x)
+ {Z\over2}\left((\partial_\mu\sigma(x))^2 +
(\partial_\mu\vec{\pi}(x))^2\right)
+ \cdots \right] \,,
\end{equation}
where Z is a renormalization constants, we obtain the effective potential
\begin{eqnarray}
V_{eff}(x) & = & {\mu^2 \over 2} (\sigma^2(x) + \vec{\pi}^2(x))
- f_{\pi} m_{\pi}^2 \sigma \nonumber\\
&   & + {3g^4 \over 8\pi^2} \left(\sigma^2(x)
+ \vec{\pi}^2(x)\right)^2\ln\left({M^2\over \Lambda^2}\right) \,.
\end{eqnarray}

The NJL model has a nontrivial vacuum, which has non-vanishing vacuum
expectation
values (VEV's) for the $\sigma$ and $\vec\pi$ fields.  These VEV's
are determined by the stationary phase conditions :
\begin{equation}
\left.{\partial V_{eff} \over \partial\sigma}\right|_{\rm vacuum}
= \left.{\partial V_{eff} \over \partial\vec{\pi}}\right|_{\rm vacuum} = 0\,,
\end{equation}
from which we get $\pi^i_{\!\!v}\,\,=\,\,<\!\!\pi^i\!\!>\,\,=\,\,0$ and
$\sigma_v\,\,=\,\,<\!\!\sigma\!\!>$
can be determined from the equation
\begin{equation}
\mu^2\sigma_v - f_\pi m_{\pi}^2 +
{3g^4\over 2\pi^2}\sigma_v^{\!3}\ln\left({M^2\over\Lambda^2}\right) = 0\,.
\end{equation}
Alternatively, we can obtained $\sigma_v$ from the mass equations for
the fields.
The masses are related to the curvatures of the effective potential
at the vacuum point; for the pion fields,
\begin{equation}
m^2_{\pi}  \equiv  \left.{\partial^2 V_{eff} \over \partial{\pi^i}^2
}\right|_{\rm vacuum}
 =  \mu^2 + {3g^4 \over 2\pi^2}\sigma_v^{\!2}
\ln\left({M^2 \over \Lambda^2}\right)\,,
\end{equation}
while for the $\sigma$ field,
\begin{equation}
m^2_{\sigma}  \equiv  \left.{\partial^2 V_{eff} \over \partial
\sigma^2}\right|_{\rm vacuum}
 =  \mu^2 + {9g^4 \over 2\pi^2}\sigma_v^{\!2}
\ln\left({M^2 \over \Lambda^2}\right)\,,
\end{equation}
where we have $\sigma_v = f_\pi$.

If we demand that the first non-vanishing term yield the proper
kinetic energy of the mesons, we set $Z = {3g^2\over 4\pi^2}
\ln\left(M^2/\Lambda^2\right) = 1$, which reduces the number of
arbitrary parameters, we obtain the regularization scheme independent
relation for the $\sigma$-mass sum rule [2],
\begin{equation}
m_\sigma^2 = 4g^2f_\pi^2 + m_\pi^2\,.
\end{equation}
In addition, from the equation of motion for the $\sigma$ field with $M_0 = 0$,
and the VEV $\sigma_v = f_\pi$, we obtain
in the soft pion limit,
a cut-off independent relation,
\begin{eqnarray}
M_0 <\bar{q}q> = - f_\pi^2 m_\pi^2\,,
\end{eqnarray}
where the value of the quark condensate $<\bar{q}q>$
provides a measure of spontanteous symmetry breaking.
\vskip 1cm
\noindent
{\large \bf 4. Discussion}

In this paper, we have calculated explicitly the effective potential
at the one-loop order in DR, and thus provided a systematic spacetime
approach, which can be readily applied to more useful extended NJL models
[4,5] in which gauge fields also play a role.

We should note that in ref. [2], several regularization schemes were used
to obtain one-loop effective potentials, which
were shown to differ substantially with respect to
renormalization scales.  This shows that the regularization methods are not
all directly applicable to chiral models like the NJL models,
and the use of those methods should be used with care.
It should be mentioned that
higher-loop corrections are generally needed to reduce the sensitivity of
renormalization scheme dependence of physical quantities.
Therefore, it is of practical importance also to provide a
calculational procedure for two- or higher-loop orders.

Finally, we should mention that in our calculation we have used the same
short-distance cut-off parameter $M$ for regularizing the divergences.
This ambiguity can be avoided if we employ the background-field method
with the symmetry of the model being enforced throughout the calculation,
such that only one cut-off parameter may be allowed.
\newpage

\noindent{\large \bf Acknowledgments}

\noindent
This research was supported by the National Science Council of the Republic
of China under Contract Nos. NSC 84-2112-M-006-012 and NSC 83-0208-M-006-057.

\end{document}